\documentclass{WileyMSP-template}

\begin{document}

\pagestyle{fancy}

\title{Quantum Machine Learning Implementations: \\Proposals and Experiments}

\maketitle


\author{Lucas Lamata*}



\begin{affiliations}
Prof. L. Lamata\\
Departamento de F\'isica At\'omica, Molecular, y Nuclear, Facultad de F\'isica, Universidad de Sevilla, Apartado 1065, 41080 Sevilla, Spain, and Instituto Carlos I de F\'isica Te\'orica y Computacional, 18071 Granada, Spain\\
Email Address: llamata@us.es

\end{affiliations}



\begin{abstract}

This article gives an overview and a perspective of recent theoretical proposals and their  experimental implementations in the field of quantum machine learning. Without an aim to being exhaustive, the article reviews specific high-impact topics such as quantum reinforcement learning, quantum autoencoders, and quantum memristors, and their experimental realizations in the platforms of quantum photonics and superconducting circuits. The field of quantum machine learning could be among the first quantum technologies producing results that are beneficial for industry and, in turn, to society. Therefore, it is necessary to push forward initial quantum implementations of this technology, in Noisy Intermediate-Scale Quantum Computers, aiming for achieving fruitful calculations in machine learning that are better than with any other current or future computing paradigm. 

\end{abstract}


\section{Introduction}

Quantum Machine Learning has emerged in the past few years as a promising field that could produce novel computing paradigms in artificial intelligence and machine learning, by means of controllable quantum devices. Recent times have witnessed an explosion of activity in this field, and several hundreds of papers specifically in this topic have been published. For reviews in quantum machine learning, see Refs.   [1-8].     Some textbooks in the field for more introductory topics are Refs. [9,10].

Even though theoretical results more related to computer science are important in the field, as they can show more easily speedup evidence via complexity-class arguments, in our view it is always important to carry out proposals for implementations, for nearer-term devices instead of full-fledged scalable quantum computers, because i) this can motivate experimental groups to push a bit further their technologies to be able to implement these proposals in the short term, and ii) there is some hope inside the quantum machine learning community that with Noisy Intermediate-Scale Quantum computers (NISQ) [11] one may be able to already achieve some kind of quantum speedup, and produce results which are useful for industry and society. In this sense, in this article we will review some theory proposals in the field of quantum machine learning, together with their experimental implementations in the quantum platforms of quantum photonics and superconducting circuits, which are two platforms that seem particularly well suited for quantum machine learning applications.

With this article we do not intend to give a thorough account of the existing literature, or even cover most of it, which is already very extensive, but to select a few theory results which have been carried out in the lab and describe them in some detail. We will focus on three proposals and their respective quantum experiments: quantum reinforcement learning, quantum autoencoders, and quantum memristors. These topics were reviewed in Ref. [6] in a more general way and with no focus on implementations. Here, we will emphasize more the experimental realizations.

In Section 2 we will describe proposals for quantum reinforcement learning and their experimental implementations with quantum photonics and superconducting circuits. In Section 3, we will review the topic of quantum autoencoders via quantum adders, and an experiment with superconducting circuits, and in Section 4 we will revisit the concept of photonic quantum memristor, as well as its implementation with a quantum photonics system. Finally, in Section 5 we will give the Conclusions.

\section{Quantum Reinforcement Learning}

The field of Quantum Reinforcement Learning deals with analyzing quantum agents that interact with their outer world, so named ``environment'' (not to be confused with the quantum environment producing decoherence), and adapt to it, following a policy in order to maximize their long-term rewards, see Figure 1. For articles reviewing the literature on quantum reinforcement learning, see, e.g., Refs. [6,8, 12-14].

\begin{figure}
  \includegraphics[width=0.6\linewidth]{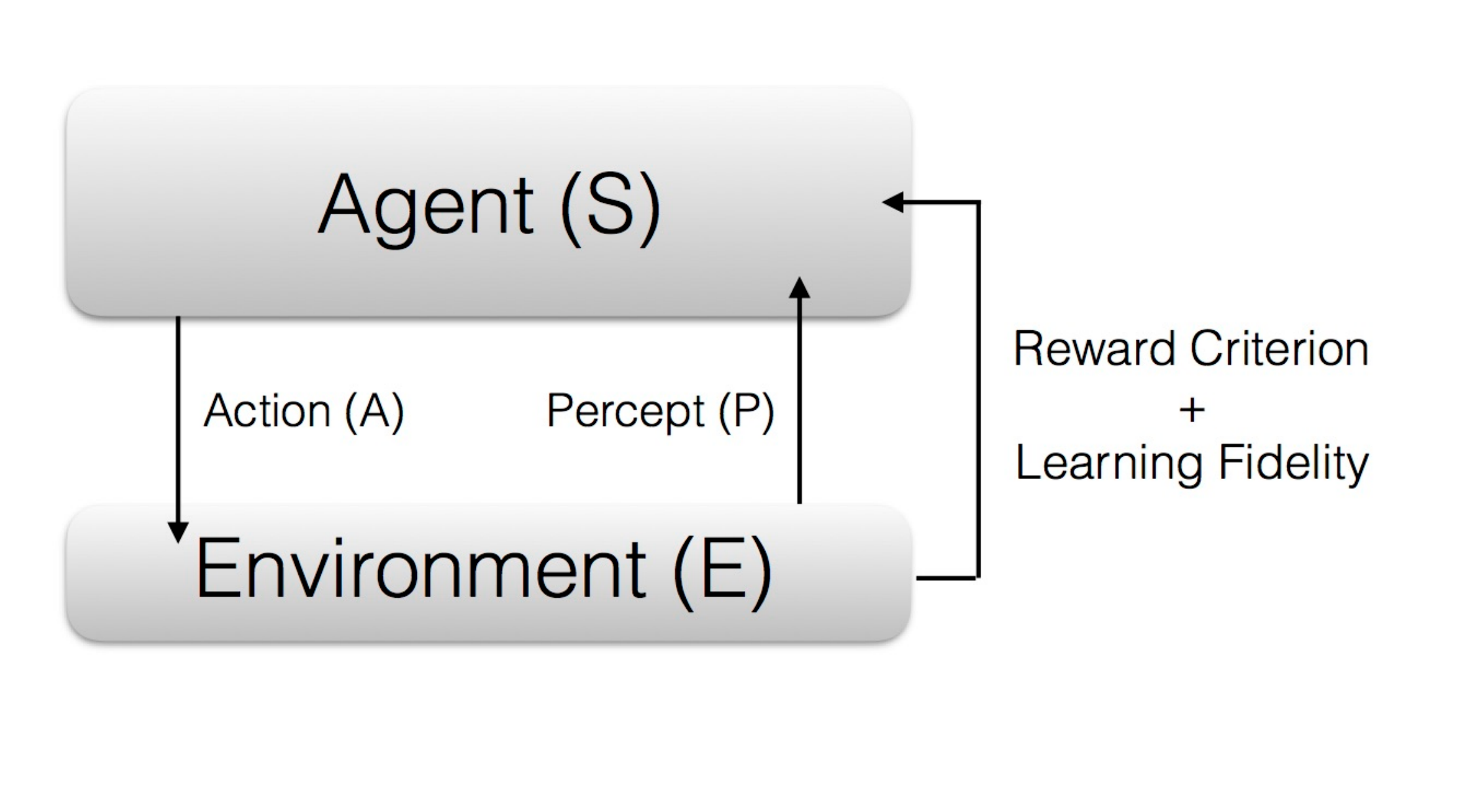}
  \caption{Scheme of Quantum Reinforcement Learning. Reproduced under terms of the CC-BY license.\textsuperscript{[Ref.  15]}. Copyright the Author, 2017. Springer Nature.}
  \label{fig:boat1}
\end{figure}

Here we will focus on a series of works [15,17-19] that proposed implementations of quantum reinforcement learning with quantum platforms such as superconducting circuits and quantum photonics, considering ingredients such as quantum and classical feedback which in some cases have been available only very recently. In these works, the aim of the quantum agent is to learn a model of the environment, namely, to learn an unknown quantum state, as an alternative protocol to more standard quantum tomography. We will first describe the original theory proposals for implementations, and subsequently review two experimental implementations that were carried out, in quantum photonics [18], in the Hefei group, and in superconducting circuits [19], with the Rigetti cloud quantum computer.

There have been other theoretical and experimental achievements in quantum reinforcement learning, in some cases with proven speedups with respect to classical protocols. Some of these works deal with a quantum agent interacting with a classical environment, with a quantum information processing via Grover search. For a comprehensive account of the related literature, see Refs.[8,6,12-14].  One highlight to mention is a pioneering experiment with quantum photonics showing a speedup that has been recently carried out [20]. 
\subsection{Theory Proposals}

One of the first proposals for experimental implementations of quantum machine learning, in this case with the platform of superconducting circuits, was put forward in Ref. [15]. In this paper, a proof-of-principle minimal scenario of quantum reinforcement learning, with a quantum agent, a quantum environment, and ways to make them interact, was introduced, in order to pave the way for experimental implementations with available superconducting circuit technology. A crucial ingredient of the analysis was the suggestion to use coherent quantum feedback, which had only very recently been achieved in the Delft group [16], as a way to enable the learning of the quantum agent. Evidence was given that with the technology available at the time several learning cycles inside the coherence time could be achieved. The goal of the agent was to acquire information from an unknown state of the environment and adapt to it, in order to achieve the same state as it (as a toy-model of quantum reinforcement learning) [15].

Subsequently, the previous proposal was extended to include several copies of the environment state, in order that the agent may interact subsequently with each of them, and progressively evolve approaching the environment state [17]. In the theoretical analysis it was shown that large quantum fidelities could be achieved at the end of the learning process. The protocol consisted on a controlled interaction between agent and environment respective states, a projective measurement, and, depending on the outcome, to apply the identity gate, or a local random unitary gate to both agent and environment (in the latter case just in order to change the measurement basis in the subsequent step). The random unitary would be nearer or farther from the identity operation depending on the previous measurement history. Namely, if one gets many 0`s, so to say, one would ``close'' the unitary range (with an exponent going to zero), while, if one gets many 1's, one would ``open'' the unitary range (by enlarging the exponent in the unitary operation). This means that, as long as we approach the environment state, we tend to close the unitary range, while, when we separate from the environment state, we open the unitary range. This is known in standard reinforcement learning as the balance between exploration and exploitation. By iterating several times this protocol, the analysis showed that the fidelities converged to near one. This would be a way to learn a quantum state in a fully automated way, with no human intervention in the process.

\subsection{Experimental Implementations}

Two experimental implementations have been carried out in the previous theory proposals. The first one in a quantum photonics platform, in the Hefei group [18], and the second one in the superconducting circuit of Rigetti cloud quantum computer [19].

\subsubsection{Implementation with quantum photonics}

The first of the implementations [18] of the previous proposal of quantum reinforcement learning in Ref. [17], was carried out in the quantum photonics Hefei laboratory, in collaboration with the theory authors, see Figure 2.

\begin{figure}[h!]
  \includegraphics[width=0.5\linewidth]{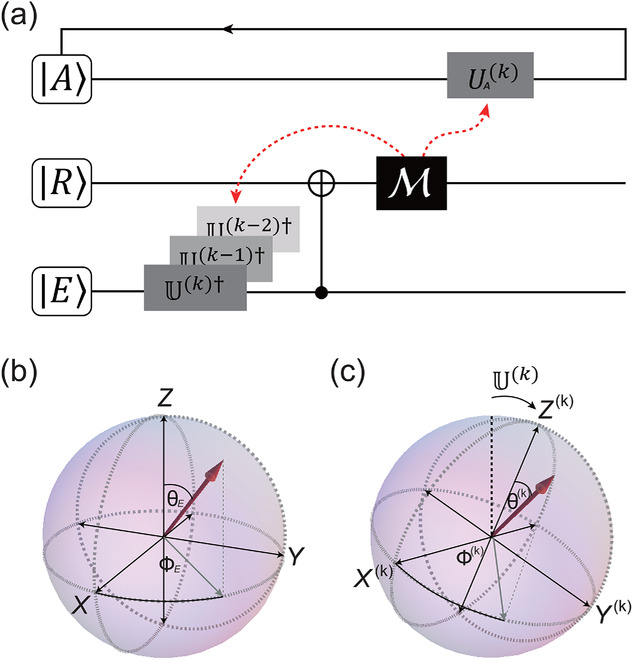}
  \caption{Quantum Reinforcement Learning protocol carried out with a quantum photonics experiment. a) Scheme of the protocol. b) environment single-qubit state as depicted in the Bloch sphere. c) the agent state approaches the environment state after a single iteration of the protocol. Reproduced with permission.\textsuperscript{[Ref. 18]} Copyright 2019, Wiley.}
  \label{fig:boat1}
\end{figure}

The experiment realized the proposal in Ref. [17] with a bulk optics quantum setup, considering a photonic quantum state of one qubit for the agent, another one for the environment, and a third one for an auxiliary register. The qubits were encoded in polarization states of the photons, and produced via Parametric Down-Conversion in a nonlinear crystal. Via quantum linear optics devices, the necessary gates were produced, and subsequently projective measurements on the obtained photonic state were carried out in each step, producing the necessary feedback on agent and environment for the next iteration, see Figure 3.

The experimental realization showed that, even with few iterations of the protocol, large fidelities could be achieved, and in some cases one could have a better performance than with just a standard quantum tomography protocol, see Figure 4. More specifically, evidence was obtained that this would take place when the quantum resources are limited, namely, in the situation that I called, in subsequent papers, ``reduced resource scenario'' [6,13,14]. This could be a complementary realm to the scaling up one, in which the aim is not to outperform classical devices via larger systems, but to have a speedup, with respect to classical or other quantum protocols, when the resources are limited, or constrained. This could also produce a benefit in industry, without the need to have a full-fledged, scalable quantum computer.

\begin{figure}[h!]
  \includegraphics[width=0.6\linewidth]{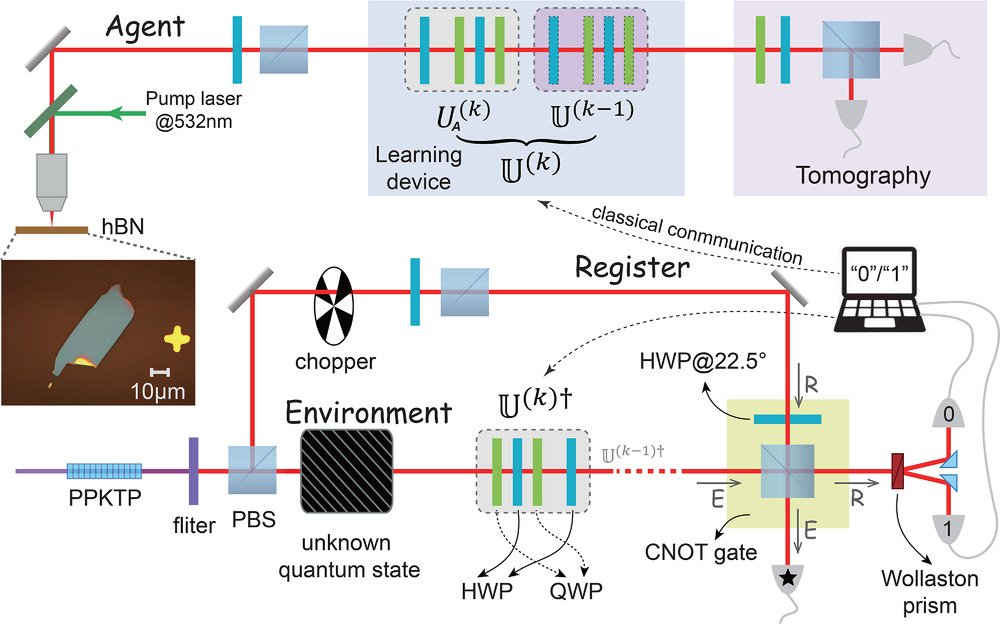}
  \caption{Experimental setup of the Quantum Reinforcement Learning protocol with quantum photonics. Reproduced with permission.\textsuperscript{[Ref. 18]} Copyright 2019, Wiley.}
  \label{fig:boat1}
\end{figure}

\subsubsection{Implementation with superconducting circuits}

A second implementation in the proposal of Ref. [17] was carried out, in this case in the superconducting circuit platform provided by Rigetti in the cloud [19]. Several instances of experimental realizations of the proposal were realized, decomposing the protocol in the available gates in the Rigetti interface at the time, and executing the Python code via connecting with Rigetti to launch the USA experiment from the University 
 of the Basque Country in Bilbao. The possibility to use feedback that the Rigetti interface provided in the 8-qubit Agave chip was an advantage for this experimental realization. Up to 6 different environment states were employed in a series of realizations of the experiment, obtaining in general large fidelities, in many cases above 0.95.
 
 \begin{figure}
  \includegraphics[width=0.6\linewidth]{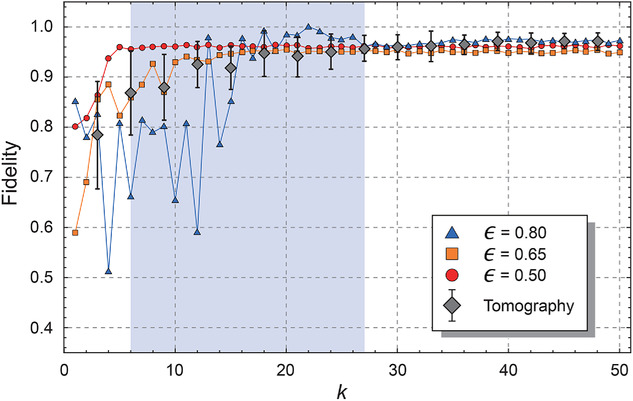}
  \caption{Experimental data of the Quantum Reinforcement Learning protocol with quantum photonics. The gray shaded region corresponds to a speedup of the quantum reinforcement learning protocol with respect to standard quantum tomography. Reproduced with permission.\textsuperscript{[Ref. 18]} Copyright 2019, Wiley.}
  \label{fig:boat1}
\end{figure}

\section{Quantum Autoencoders}

Inside the field of Quantum Machine Learning, the topic of Quantum Autoencoders aims at employing quantum resources in a more efficient way, by discarding unnecessary parts of the quantum device [21,22]. The idea is to encode the quantum states under consideration in the minimum number of quantum bits necessary, that in general will be smaller than the original Hilbert space under study. For this, one would employ a parameterized quantum circuit with a smaller amount of qubits inside than in the outer parts, and try to match the final outcome of the quantum circuit with the input state via changing the parameters of the quantum unitary gates. If this quantum-classical procedure finally converges, then one can discard the decoding, final part of the circuit and keep just the encoding part, which maps the original quantum states onto a smaller amount of qubits.

Here we will describe one of the existing proposals for quantum autoencoders, that employs the concept of a ``quantum adder'' [23]. Some years ago it was proven that no unitary gate exists capable of creating a quantum superposition of two unknown quantum states [24,25]. However, it was also shown that approximate adding quantum operations exist, which create superpositions of quantum states to a certain fidelity and/or probability [24,26]. Therefore, it was just natural to consider this kind of operations, shown to be able to optimize with genetic algorithms [26], as a toolbox for implementing quantum autoencoders [23]. The reason is that, if one applies a quantum adder to, say, two unknown quantum states, one could encode the two quantum states onto a single quantum state, namely, the approximate quantum superposition. In this way, compression of the quantum information from two states to a single one would be achieved. The fidelity and probability for this task, which would in general be smaller than one, would depend on the specific initial quantum states and the approximate quantum adder employed.

We will also describe an experimental implementation of the previous proposal with superconducting circuits in the Rigetti cloud quantum computer [27].

There is another experimental result of a quantum autoencoder in the literature, involving a quantum photonics platform [28].

\subsection{Theory Proposal}

The proposal we will describe here deals with implementing a quantum autoencoder via quantum adders optimized with genetic algorithms [23]. To this aim, firstly a result dealing with the implementation of quantum adders optimized with genetic algorithms in the IBM Quantum Experience cloud quantum computer was obtained [26]. The genetic algorithm employed a series of quantum gates as the genetic code, which were numerically optimized, via this machine learning algorithm, in order to maximize some figure of merit, such as, e.g., fidelity. The gates were randomly combined in ``genes'', mutated, and different genes were mixed. The most successful combinations were kept, and the rest discarded, and the protocol was iterated until convergence, see Figure 5. This was later on employed as the basis for the approximate quantum adders implementing the quantum autoencoders [23]. Once the quantum adders were achieved, the definition of the quantum autoencoders was straightforward, see Figure 6. A bonus of this approach is that, in addition to having the final quantum information encoded in a smaller quantum subspace, there is a possibility to decode the process via keeping part of the initial quantum state information in ``memory qubits'', which can then, via the inverse unitary operation to the quantum adder, help to retrieve the initial state. Of course, given that the quantum adder operation cannot work perfectly for every quantum state, the complete protocol of encoding and decoding cannot have fidelity one for every quantum state. Nonetheless, there could be situations, such as in quantum satellites or quantum drones, in which an encoding of the quantum information onto a smaller amount of qubits could be beneficial for saving valuable resources.

\begin{figure}
  \includegraphics[width=0.7\linewidth]{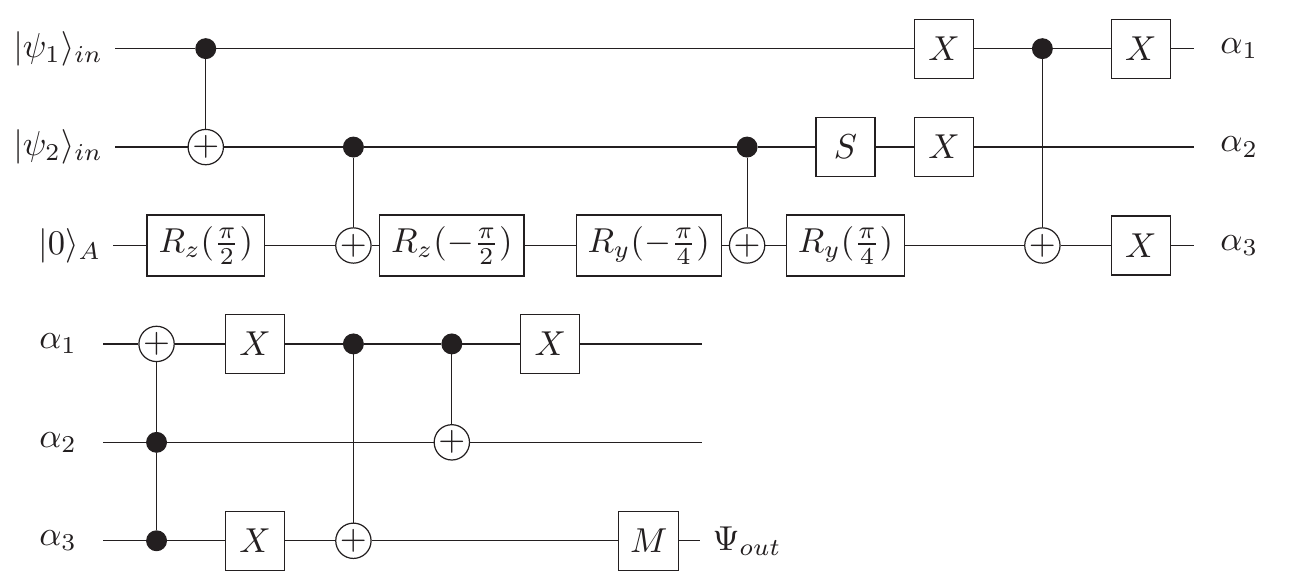}
  \caption{Gate circuit for the basis quantum adder. Reproduced with permission.\textsuperscript{[Ref. 27]} Copyright 2019, Wiley.}
  \label{fig:boat1}
\end{figure}

In the same article another proposal for implementing quantum autoencoders was introduced, namely, instead of employing quantum adders optimized with genetic algorithms, the idea is to directly optimize the quantum autoencoders with genetic algorithms [23]. This could avoid local minima with respect to alternative proposals for quantum autoencoders in the literature, employing other optimization subroutines such as gradient descent.

The proposal for quantum autoencoders with quantum adders optimized with genetic algorithms was carried out in the Rigetti cloud quantum computer [27], as we describe next.

\subsection{Experimental Implementation with Superconducting Circuits}

The previous proposal was carried out in the superconducting circuit cloud quantum computer by Rigetti, employing up to three qubits [27]. The fidelities achieved in the experiment agreed well with the theoretical predictions of the proposal, showing the feasibility to implement quantum autoencoders in superconducting circuits via this approach.

\begin{figure}
  \includegraphics[width=0.5\linewidth]{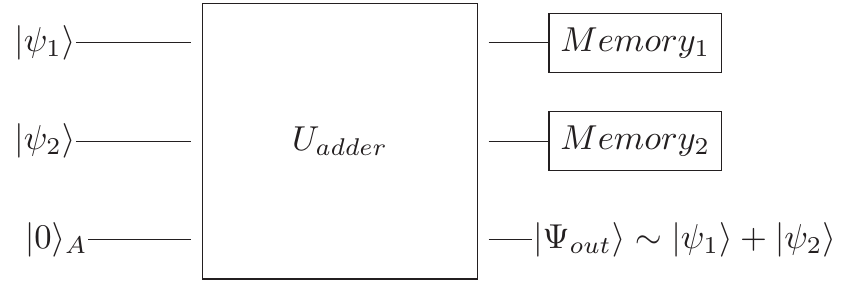}
  \caption{Scheme for a quantum autoencoder based on quantum adders. Reproduced with permission.\textsuperscript{[Ref. 27]} Copyright 2019, Wiley.}
  \label{fig:boat1}
\end{figure}

In this experiment, two different quantum adders optimized with genetic algorithms were considered. Namely, the basis quantum adder and the gate-limited quantum adder [26], see Figure 7. The basis quantum adder adds perfectly well the computational basis states, and adds their superpositions with smaller fidelity, but it employs a larger amount of quantum gates than the gate-limited quantum adder. As expected, the fidelity in the latter is smaller, as the optimization is not as good with more constrained resources. However, the accumulated gate error will be smaller than the protocol with larger amount of gates. Therefore, with current NISQ [11] quantum computers one will have to look for the optimal regime that employs enough gates to reduce the ideal algorithmic error, but not to use too many gates such that the accumulated gate error is not too large.

\begin{figure}
  \includegraphics[width=0.6\linewidth]{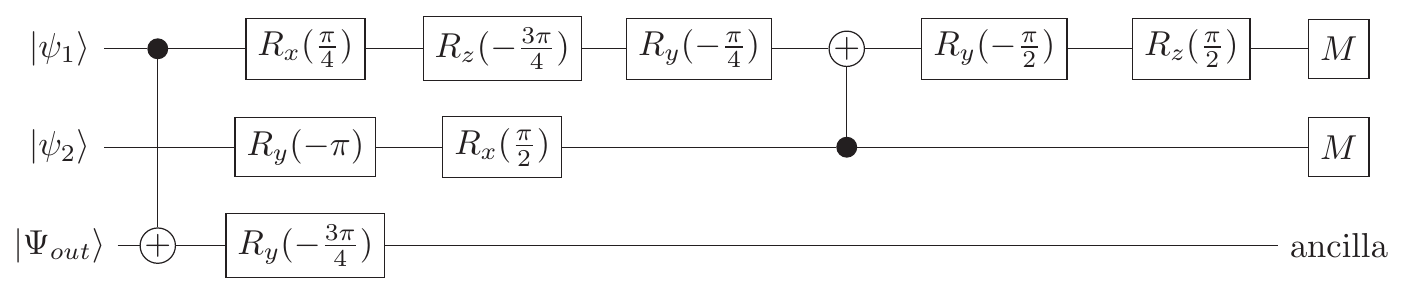}
  \caption{Decoder circuit for the quantum autoencoder based on the gate-limited quantum adder. Reproduced with permission.\textsuperscript{[Ref. 27]} Copyright 2019, Wiley.}
  \label{fig:boat1}
\end{figure}

\section{Quantum Memristors}

The concept of quantum memristors [29] is somehow connected to the one of quantum reinforcement learning already described in Section 2, as it also considers feedback in quantum systems, but it is also different in the sense that a quantum memristor would be equipped with a weak measurement, which keeps part of the quantum coherence and entanglement of the network of quantum memristors.  These quantum devices are also nonlinear at the wavefunction level, and as such can provide, in principle, more efficient learning than with just purely unitary systems, as well as quantum simulations of non-Markovian quantum systems. Quantum memristors are also systems with hysteresis, and, in this sense, memory, as they keep track of the past history of states [29].

Here we will describe a proposal for implementation of quantum memristors with quantum photonics [30], and a pioneering experimental realization in the Vienna quantum photonics lab [31,32].

\subsection{Theory Proposals}

A quantum memristor was defined in Ref. [29] as an elementary nonlinear quantum system composed of a qubit, weak measurements on its quantum state, and feedback onto the same quantum memristor or other ones. This simple concept allows one for a rich implementation of quantum systems with memory, in the sense of hysteresis, that keep track of past states of the quantum memristor and employ those past states to produce the new updates of the system. Among the quantum platforms that were proposed for implementation of this concept, quantum photonics [30] and superconducting circuits [33] appear as prominent ones, due to the high-degree of controllability, fabrication, and ease of connecting them into networks. Further proposals in this area have also been developed [34].

In Ref. [30] a proposal for a photonic quantum memristor was introduced and analyzed. This concept relied on a tunable beam splitter, where the output of one of the ports was connected via feedback onto the same beam splitter in order to modify its reflectivity, and this produced a nonlinear behaviour of the quantum photonic state, as well as a hysteresis loop, see Figure 8. This basic building block was shown to have quantum character, in the sense of quantum coherence, and would be able to be connected with other quantum memristors in a global network involving entanglement as well as nonlinear behaviour.

\begin{figure}
  \includegraphics[width=0.4\linewidth]{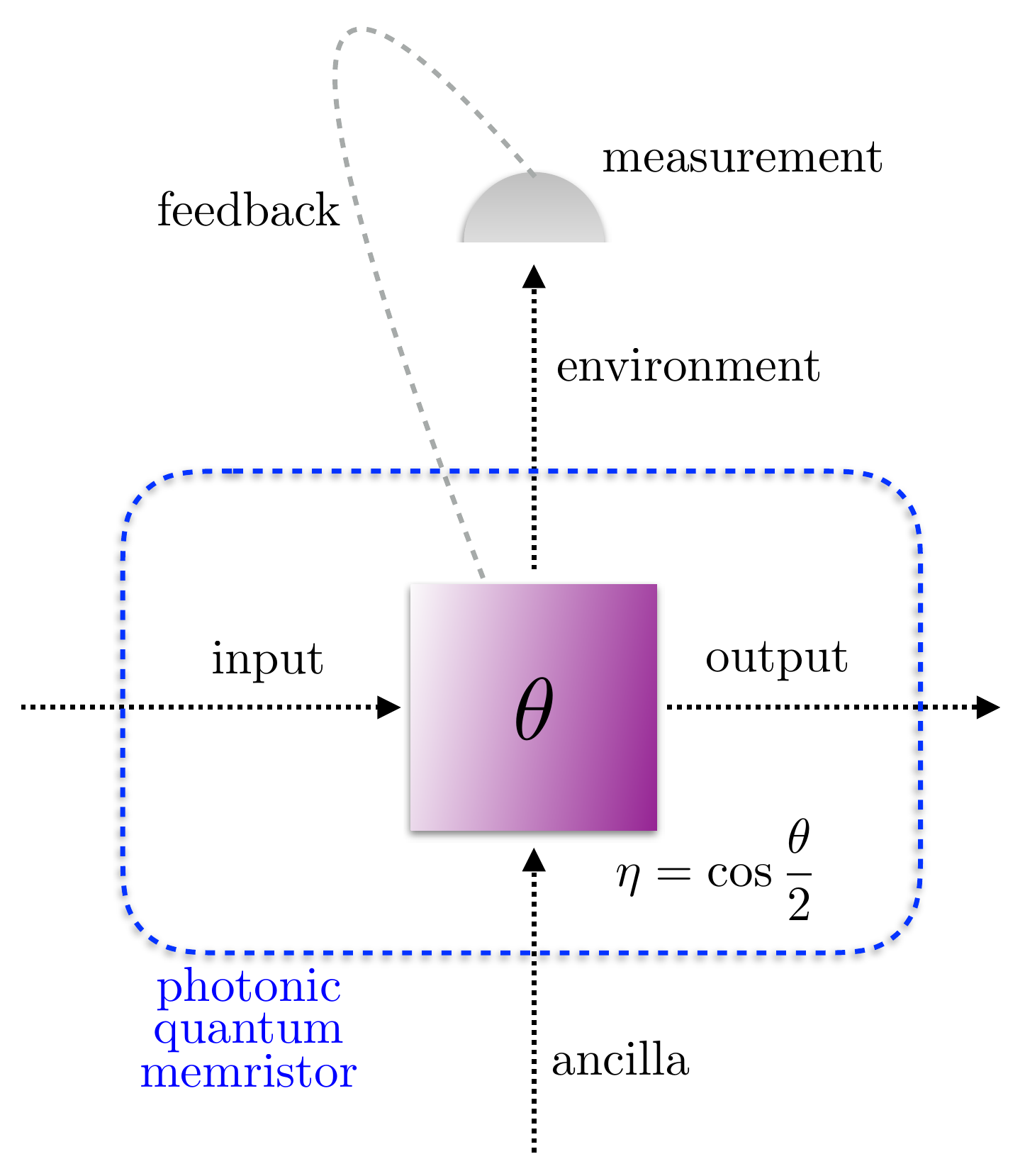}
  \caption{Scheme for a photonic quantum memristor. Reproduced under terms of the CC-BY license.\textsuperscript{[Ref. 30]} Copyright the Authors, 2018, AIP.}
  \label{fig:boat1}
\end{figure}

Even though originally speedup considerations were not taken into account in these proposals, in the photonic quantum experiment carried out in 2022 in the Vienna lab [31], in collaboration with Milan research groups, some evidence for quantum speedup with these devices was shown. We describe next the experimental implementation in Vienna.

\subsection{Experimental Implementation with Quantum Photonics}

A photonic quantum memristor was experimentally realized in 2022 in the Vienna quantum photonics lab in collaboration with research groups in Milan [31,32], based on the theory proposal of Ref. [30]. The photonic quantum memristor was built on a photonic chip with a glass-made, laser-written substrate. The device is based on a Mach-Zehnder interferometer, where the reflectivity is modified by a phase-shifter in one of the arms. The experimental photonic quantum memristor showed typical hysteretic behaviour in agreement with the theory predictions, as shown in Figure 9.

Interestingly, in this paper the authors of the experiment also carried out numerical simulations including three quantum memristors inside a quantum neural network of reservoir computing, and compared it to other previous purely classical results. Although it is hard to rigorously and sensibly compare the outcomes of different neural networks, either classical or quantum, apparently they obtained a speedup with their classically-simulated network of quantum memristors, with respect to similar purely classical previous calculations. Even though the classification problems considered were not fully equivalent in these two cases, this could point to a gain by employing quantum memristors inside photonic networks. Furthermore, given that the good performance that they achieved was based on a classical simulation of just three quantum memristors, I wonder whether this could produce a novel purely classical machine learning algorithm for classification, along the lines of quantum-inspired classical algorithms [35], incorporating, inside a neural network, classical simulations of quantum memristors, instead of the real physical ones, at least for small sizes. Perhaps the good performance could be a consequence of the nonlinear and coherent behaviour of each simulated quantum memristor, and their entanglement, or perhaps it could be just an artifact of the specific cases analyzed. In any case, both the experimental quantum memristors, and the simulated quantum memristors inside quantum neural networks and reservoir computing, seem to me as promising avenues to be further explored, which could give rise perhaps to novel and useful computing paradigms.

\begin{figure}
  \includegraphics[width=0.7\linewidth]{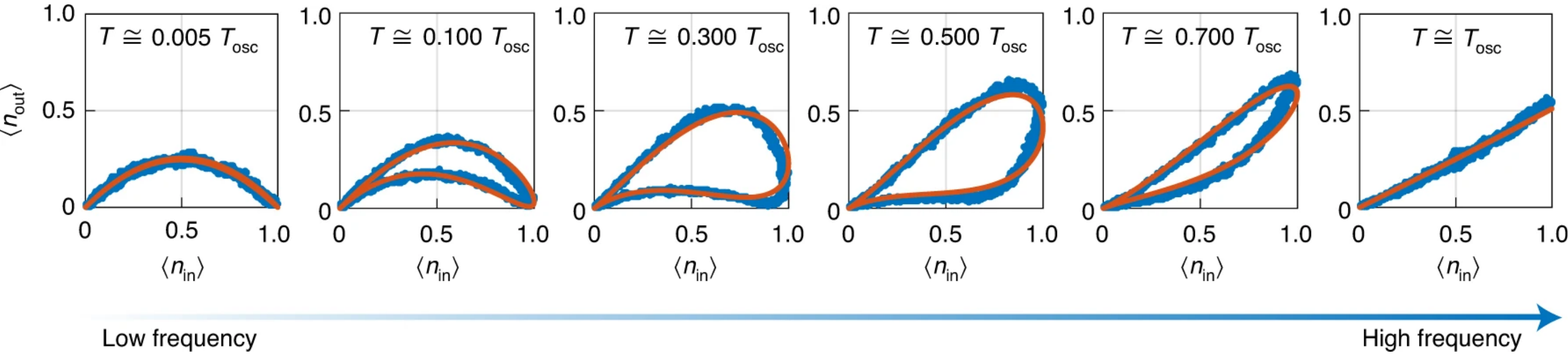}
  \caption{Hysteresis loops in an experimental implementation of a photonic quantum memristor, where experimental data are blue lines and simulated dynamics are red lines. Reproduced under terms of the CC-BY license.\textsuperscript{[Ref. 31]} Copyright the Authors, 2022, Springer Nature.}
  \label{fig:boat1}
\end{figure}

\section{Conclusion}

Quantum Machine Learning is an enticing research field that could provide a novel computing paradigm for more efficient machine learning calculations. In the near term, it is important to carry out proof-of-principle quantum machine learning experiments that may already show some evidence of speedup with respect to classical algorithms. In this overview and perspective, we have described three theoretical results inside quantum machine learning, and their experimental implementations with the controllable quantum platforms of quantum photonics and superconducting circuits. We hope that this, in our view, interesting research, may further motivate other scientists to push forward the field of quantum machine learning implementations, both at the theory and experiment levels. It is amazing that, since we edited a Special Issue on Quantum Machine Learning and Bioinspired Quantum Technologies in 2019 in Advanced Quantum Technologies [36], the field has grown beyond expectations, and we hope that it continues to do so.


\medskip
\textbf{Acknowledgements} \par 
I acknowledge funding by the Junta de Andaluc\'ia (P20-00617 and US-1380840) and by the Spanish Ministry of Science, Innovation, and Universities under grant Nos. PID2019-104002GB-C21 and PID2019-104002GB- C22.

\medskip
\textbf{Conflicts of Interest} \par

The author declares no conflicts of interest.
\medskip

%

\textbf{References}\\

1	Biamonte, J., Wittek P, Pancotti N, Rebentrost P, Wiebe N and Lloyd S 2017 Quantum machine learning Nature 549 074001 \\

2     Dunjko V and Briegel H J 2018 Machine learning \& artificial intelligence in the quantum domain: a review of recent progress Rep. Prog. Phys. 81 074001\\

3 Schuld M, Sinayskiy I and Petruccione F 2015 An introduction to quantum machine learning Contemp. Phys. 56 172\\

4 Schuld M, Sinayskiy I and Petruccione F 2014 The quest for a quantum neural network Quantum Inf. Process. 13 2567\\

5 Dunjko V and Wittek P 2020 A non-review of Quantum Machine Learning: trends and explorations Quantum 4 32\\

6 Lamata L 2020 Quantum machine learning and quantum biomimetics: A perspective Mach. Learn.: Sci. Technol. 1 033002\\

7 Martín-Guerrero J and Lamata L 2022 Quantum Machine Learning: A tutorial Neurocomputing 470 457\\   

8 Melnikov A, Kordzanganeh M, Alodjants A and Lee R-K 2023 Quantum machine learning: from physics to software engineering, Adv. Phys. X 8 2165452\\

9	Wittek P 2014 Quantum Machine Learning (New York: Academic Press) \\

10        Schuld M and Petruccione F 2018 Supervised Learning With Quantum Computers (Berlin: Springer)\\

11 Preskill J 2018 Quantum Computing in the NISQ era and beyond Quantum 2 79

12 Dunjko V, Taylor J, and Briegel H 2017 Advances in quantum reinforcement learning IEEE International Conference on Systems, Man, and Cybernetics (SMC), 282\\

13 Lamata L 2021 Quantum Reinforcement Learning with Quantum Photonics, Photonics 8 33 \\

14 Mart\'in-Guerrero J and Lamata L 2021 Reinforcement Learning and Physics  Appl. Sci. 11 8589\\

15 Lamata L 2017 Basic protocols in quantum reinforcement learning with superconducting circuits Sci. Rep. 7 1609

16 Ristè D and DiCarlo L, 2016 Digital Feedback Control, Superconducting Devices in Quantum Optics pp 187 (Springer)

17 Albarrán-Arriagada F, Retamal J C, Solano E and Lamata L 2018 Measurement-based adaptation protocol with quantum reinforcement learning Phys. Rev. A 98 042315

18 Yu S, Albarr\'an-Arriagada F, Retamal J C, Wang Y-T, Liu W, Ke Z-J, 
    Meng Y, Li Z-P, Tang J-S, Solano E, Lamata L, Li C-F, and Guo G-C 2019 Reconstruction of a Photonic Qubit State with Reinforcement Learning Adv. Quantum Technol. 2 1800074

19 Olivares-S\'anchez J, Casanova J, Solano E, and Lamata L 2020 Measurement-based adaptation protocol with quantum reinforcement learning in a Rigetti quantum computer Quantum Rep. 2 293

20 Saggio V, Asenbeck B E, Hamann A, Str\"omberg T, Schiansky P, Dunjko V, Friis N, Harris N C, Hochberg M, Englund D, W\"olk S, Briegel H J and Walther P 2021 Experimental quantum speed-up in reinforcement learning agents Nature 591 229

21 Romero J, Olson J P and Aspuru-Guzik A 2017 Quantum autoencoders for efficient compression of quantum data Quantum Sci. Technol. 2 045001

22 Wan K H, Dahlsten O, Kristjánsson H, Gardner R and Kim M S 2017 Quantum generalisation of feedforward neural networks npj Quantum Inform. 3 36

23 Lamata L, Alvarez-Rodriguez U, Martín-Guerrero J D, Sanz M and Solano E 2018 Quantum autoencoders via quantum adders with genetic algorithms Quantum Sci. Technol. 4 014007

24 Alvarez-Rodriguez U, Sanz M, Lamata L and Solano E 2015 The forbidden quantum adder Sci. Rep. 5 11983

25 Oszmaniec M, Grudka A, Horodecki M and Wójcik A 2016 Creating a superposition of unknown quantum states Phys. Rev. Lett. 116 110403

26 Li R, Alvarez-Rodriguez U, Lamata L and Solano E 2017 Approximate quantum adders with genetic algorithms: An IBM quantum experience Quantum Meas. Quantum Metrol. 4 1

27 Ding Y, Lamata L, Sanz M, Chen X and Solano E 2019 Experimental implementation of a quantum autoencoder via quantum adders Adv. Quantum Technol. 2 1800065

28 Pepper A, Tischler N and Pryde G J 2019 Experimental realization of a quantum autoencoder: The compression of Qutrits via machine learning Phys. Rev. Lett. 122 060501

29 Pfeiffer P, Egusquiza I L, Di Ventra M, Sanz M and Solano E 2016 Quantum memristors Sci. Rep. 6 9507

30 Sanz M, Lamata L and Solano E 2018 Invited article: Quantum memristors in quantum photonics APL Phot. 3 080801

31 Spagnolo M, Morris J, Piacentini S, Antesberger M, Massa F, Crespi A, Ceccarelli F, Osellame R and Walther P 2022 Experimental photonic quantum memristor Nature Phot. 16 318

32 Lamata L 2022 Memristors go quantum Nature Phot. 16 265

33 Salmilehto J, Deppe F, Di Ventra M, Sanz M and Solano E 2017 Quantum memristors with superconducting circuits Sci. Rep. 7 42044

34 Shevchenko S N, Pershin Y V and Nori F 2016 Qubit-based memcapacitors and meminductors Phys. Rev. Applied 6 014006

35 Tang E 2019 A quantum-inspired classical algorithm for recommendation systems STOC 2019: Proceedings of the 51st Annual ACM SIGACT Symposium on Theory of Computing 217

36 Lamata L, Sanz M and Solano E 2019 Quantum machine learning and bioinspired quantum technologies Adv. Quantum Technol. 2 1900075



\begin{figure}[h!]
  \includegraphics[width=40mm]{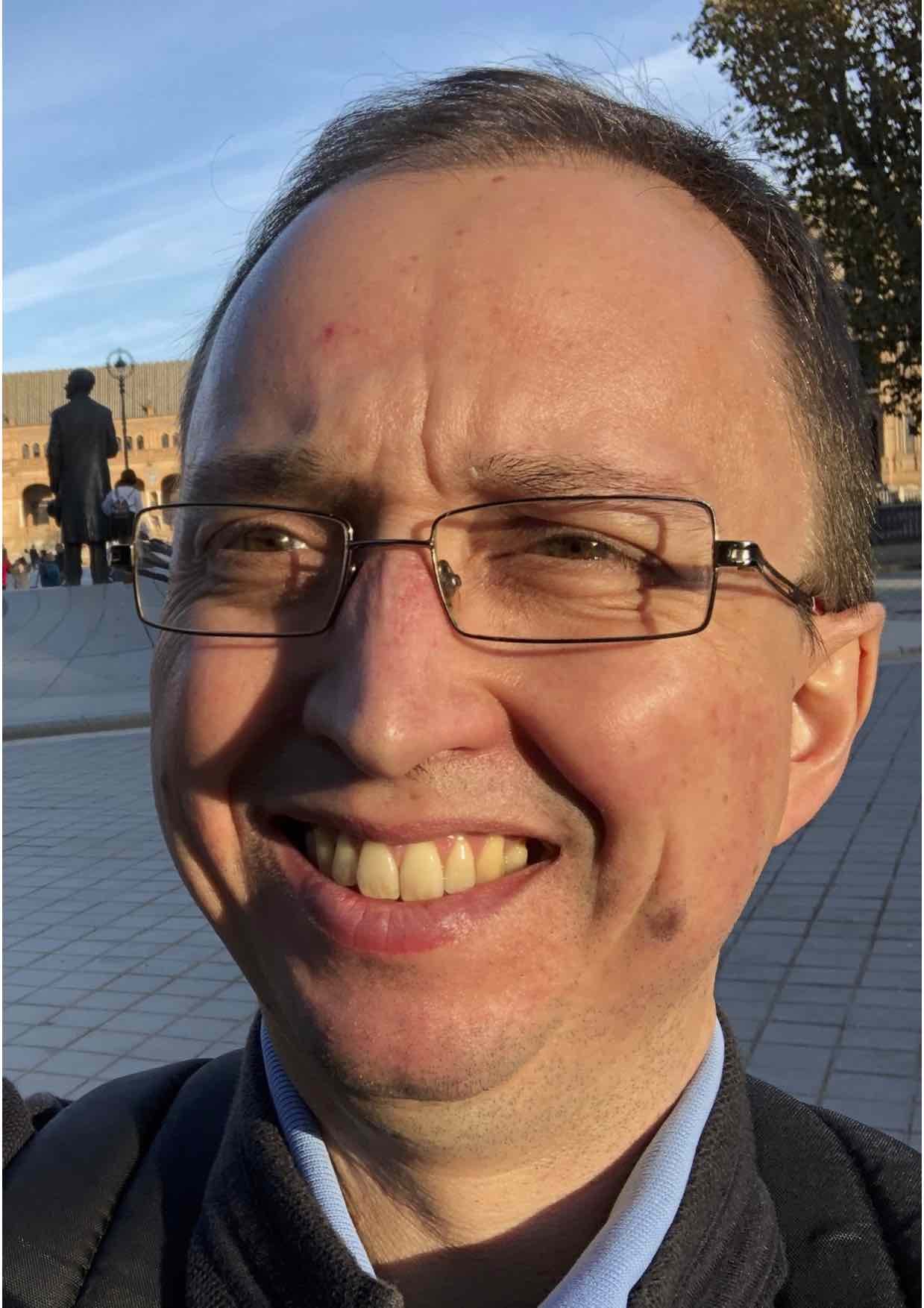}
  \caption*{Biography}
\end{figure}

Prof. Lucas Lamata is an Associate Professor (Profesor Titular de Universidad) of Theoretical Physics at the Departamento de F\'isica At\'omica, Molecular y Nuclear, Facultad de F\'isica, Universidad de Sevilla, Spain. Before working in Sevilla, he was a Marie Curie postdoctoral fellow, a Ram\'on y Cajal Fellow, and a Staff Researcher at the University of the Basque Country. Before that, he was a Humboldt Fellow and a Max Planck postdoctoral fellow for 3 and a half years at the Max Planck Institute for Quantum Optics in Garching, Germany. Previously, he carried out his PhD at CSIC, Madrid.


\end{document}